# Computational identification of ketone metabolism as a key regulator of sleep stability and circadian dynamics via real-time metabolic profiling

Hao Huang[1], Kaijing Xu[1] and Michael Lardelli[1]

## Abstract

Metabolism plays a crucial role in sleep regulation, yet its effects are challenging to track in real time. This study introduces a machine learning-based framework to analyze sleep patterns and identify how metabolic changes influence sleep at specific time points.

We first established that sleep periods in *Drosophila melanogaster* function independently, with no causal relationship between different sleep episodes. Using gradient boosting models and explainable artificial intelligence techniques, we quantified the influence of time-dependent sleep features. Causal inference and autocorrelation analyses further confirmed that sleep states at different times are statistically independent, providing a robust foundation for exploring metabolic effects on sleep.

Applying this framework to flies with altered monocarboxylate transporter 2 expression, we found that changes in ketone transport modified sleep stability and disrupted transitions between day and night sleep. In an Alzheimer's disease model, metabolic interventions such as β-hydroxybutyrate supplementation and intermittent fasting selectively influenced the timing of day-to-night transitions rather than uniformly altering sleep duration. Autoencoder-based similarity scoring and wavelet analysis reinforced that metabolic effects on sleep were highly time-dependent.

This study presents a novel approach to studying sleep-metabolism interactions, revealing that metabolic states exert their strongest influence at distinct time points, shaping sleep stability and circadian transitions.

Author affiliations:

1 Faculty of Sciences, Engineering and Technology, University of Adelaide, North Terrace, Adelaide, 5005, SA, Australia.

Correspondence to: Hao Huang

Full address: Gate8 Victoria Dr, The University of Adelaide Adelaide, North Terrace, SA5005, Australia

E-mail: hao.huang01@adelaide.edu.au

# Introduction

In animal physiology, sleep is not only essential for memory consolidation and recovery but is also closely linked to metabolic regulation. Recent studies highlight the critical role of metabolic fluctuations in sleep regulation, influencing energy balance, foraging behavior, and circadian synchronization. For instance, specific genes such as Ade2, expressed in adipose tissue, have been shown to regulate sleep duration and energy storage, while hunger suppresses sleep to promote foraging[1]. Additionally, factors such as geographical environment, temperature variations, and body size shape sleep patterns and metabolic adaptations[2]. These findings suggest that sleep is not merely a neurological necessity but also a key mechanism for maintaining energy homeostasis and survival strategies across diverse ecological conditions.

In recent years, researchers have increasingly focused on the dynamic fluctuations of metabolites during sleep and their role in sleep regulation. Metabolites such as glucose, lipids, and neurotransmitters oscillate with the sleep-wake cycle, reflecting energy metabolism status and potentially influencing sleep quality and duration. For instance, studies have shown a significant correlation between free glucose levels and sleep duration, while glycogen and triglyceride storage are linked to hunger toleranc[3–6]. Moreover, investigations at the neuronal level have revealed the role of specific genes in metabolite sensing. In Drosophila, for example, LHLK neurons suppress sleep under starvation by modulating the insulin signaling pathway[7]. These findings drive further exploration into how metabolic fluctuations shape sleep quality, architecture, and sleep disorders, positioning the temporal dynamics of metabolite concentrations as a critical direction in sleep research.

In many sleep-related disorders, fluctuations in metabolite concentrations are not only integral to the underlying pathology but also provide a foundation for developing new therapeutic strategies. For example, patients with obstructive sleep apnea (OSA) often exhibit abnormal glucose metabolism, insulin resistance, and lipid dysregulation, which are closely linked to intermittent hypoxia and sleep fragmentation[8–10]. Interventions targeting these metabolic alterations—such as dietary optimization, physical activity, and pharmacological or behavioral therapies aimed at improving sleep quality—have become central to managing sleep-related metabolic disorders[11–14]. Moreover, studies indicate that modulating circadian clock genes and metabolic signaling pathways can partially restore normal sleep and metabolic rhythms, offering new possibilities for personalized medicine and precision therapies[15–18]. Therefore, a deeper understanding of the dynamic fluctuations in metabolite concentrations not only helps elucidate the mechanisms underlying sleep disorders but also lays the groundwork for developing more effective interventions and treatment strategies.

Metabolic fluctuations are a key force shaping sleep in organisms. However, effective methods to track temporal changes in metabolites remain lacking. For most metabolites, collection and detection without disrupting sleep is nearly impossible. This study addresses this challenge through an indirect approach. Using the GS-GAL4 *Drosophila* model or dietary supplementation to decrease or increase specific metabolites, we generated fly models exhibiting altered sleep patterns[19]. Our findings reveal that sleep homeostasis and circadian rhythms in *Drosophila* are relatively independent within a 24-hour cycle, with no significant correlation between non-consecutive hours within a day or between short-term diurnal rhythms. This temporal independence forms the basis of our analytical framework, allowing us to treat sleep patterns at each hour as discrete features rather than a continuous trend. Using machine learning, we compared sleep characteristics under different metabolic conditions at each hourly time point and assigned an importance score to reflect their contribution to sleep pattern changes. By tracking changes in this score over time, we can pinpoint when the corresponding metabolite has the greatest impact.

This approach identified distinct temporal windows in which ketone bodies—fatty acid-derived metabolites previously implicated in sleep regulation—modulated sleep homeostasis and circadian stability in this research[20]. Furthermore, we applied this framework to an Alzheimer's disease (AD) *Drosophila* model, uncovering the temporal profile of ketone body intervention and its functional role in mitigating sleep disruption.

# Materials and methods

## *Drosophila* Stocks

The following *Drosophila* stocks were used in this study, maintained under a 12-hour light/12-hour dark cycle at 60% relative humidity: Tim-GAL4 (BDSC #7126), Elav-GS-GAL4 (BDSC #43642), UAS-APP+BACE1 (BDSC #33797), UAS-sln RNAi (VDRC #109464). These stocks were either generated, maintained, or provided by Ms. Louise O'Keefe for this study. For all Drosophila strains carrying Elav-GS-Gal4, transgene expression was induced by diluting the 10 mM RU486 stock solution (Sigma Aldrich, Cat. #:M8046) in 70% ethanol to the desired working concentration, mixed in fly food for use. BHB (2 mM, Sigma Aldrich, Cat. #: H6501) were administered by transferring larvae to supplemented food for and treat at least 10 days after emergence. Intermittent fasting (IF) refers to 3 times a week, each time for 16-24 hours on 0.5% agar.

## Drosophila Activity Monitoring

The Drosophila Activity Monitoring (DAM) assay was conducted at 25°C and 60% relative humidity using the DAM system (Trikinetics Inc.) to record locomotor behavior. Each experimental group comprised 32 individual flies, each housed in a separate channel. Flies were continuously monitored for seven days under a 12:12 light-dark cycle, with a separate cohort maintained in complete darkness for circadian rhythm analysis.

During the assay, flies were provided a diet of 6% sucrose in 0.5% agar to prevent egg-laying by females. Sleep was defined as any period of complete inactivity lasting five minutes or longer. Sleep patterns were analyzed using the Rethomics package (0.3.1) in R for precise quantification of sleep metrics[21]. Statistical comparisons between groups were performed using the Wilcoxon rank-sum test, as outlined in the Rethomics documentation.

## Data Processing

To accommodate different analytical needs, we employed two normalization approaches. For daily activity data processing, we applied individual activity ratio normalization by calculating the total activity of each individual over the entire experimental period and normalizing each time point by this total:

$$normalized\_activity(i, t) = \frac{activity(i, t)}{\sum activity(i, :)}$$

where normalized_activity(i, t) represents the normalized activity of individual iii at time ttt, activity(i, t) is the raw activity count, and sum(activity(i, :)) denotes the total activity of individual iii across the experiment. To prevent division by zero errors, individuals with a total activity count of zero were assigned a normalized activity value of zero across all time points. This normalization method ensures comparability of relative activity levels across individuals while eliminating inter-individual differences in overall activity, thereby reducing potential biases in causal analysis.

For fixed-window time series data, we applied Z-score normalization to ensure that data from different time windows were on the same scale. The Z-score normalization was computed as follows:

$$z\_score(i, t) = \frac{activity(i, t) - mean(activity(:, t))}{std(activity(:, t))}$$

where activity(i, t) represents the activity of individual iii at time ttt, and mean(activity(:, t)) and std(activity(:, t)) correspond to the mean and standard deviation of all individuals' activity at time ttt, respectively. This transformation ensures a zero mean and unit variance, improving comparability of activity levels across different time points and mitigating potential heteroskedasticity in the data. Z-score normalization is particularly suited for cross-window analyses, such as causal inference over two-hour intervals, as it accounts for variations in individual and population-wide activity levels across different times of the day.

For temporal aggregation, we computed mean activity levels using both daily and two-hour windows. For daily activity data, we divided the seven-day period into individual days and computed the mean activity per day. For hourly window data, we used 30-minute intervals and constructed 12 two-hour bins per day, providing a higher temporal resolution for activity pattern analysis.

# DoWhy- Causal Relationships Analysis

To infer causal relationships, we applied DoWhy for causal inference, using linear regression (backdoor.linear_regression) to estimate causal effects between time points[22,23]. Specifically, we constructed a causal model in which each time point served as either a potential outcome (effect) or predictor (cause) and evaluated direct causal relationships between time points without introducing additional confounding factors.

For instance, when assessing causal effects between Day 1 → Day 2 or 00:00 → 02:00, we defined the causal relationship using the model:

$$Y = \beta X + \epsilon$$

where Y represents the dependent variable corresponding to the activity level at the effect time point, X represents the independent variable corresponding to the activity level at the cause time point, β is the estimated causal effect, and ε is the error term. The causal effect β was estimated using the covariance formula:

$$\beta = \frac{Cov(X,Y)}{Var(X)}$$

where Cov(X, Y) denotes the covariance between X and Y, and Var(X) represents the variance of X.

The regression coefficient from this model served as the causal effect estimate, with a 95% confidence interval (CI) calculated to assess its precision. Standard errors were derived from the confidence interval bounds, and the z-statistic was computed as:

$$z\_statistic = \frac{causal\_effect}{std\_err}$$

while the corresponding two-tailed p-value was determined based on the standard normal distribution:

$$p\_value = 2 \times (1 - norm\_cdf(|\ z\_score\ |))$$

where norm_cdf represents the cumulative distribution function (CDF) of the standard normal distribution.

## Autocorrelation Function

This study employed the autocorrelation function (ACF) to analyze Drosophila activity data, aiming to explore the internal structure and rhythmic characteristics of the time series[24]. ACF is primarily used to evaluate the similarity of activity levels at different time lags, revealing the periodicity and stability of behavioral patterns. The autocorrelation function measures the correlation of a time series with itself at different lag points and is computed using the following formula:

$$ACF(k) = \frac{Cov(X_t, X_{t-k})}{Var(X_t)}$$

where ACF(k) represents the autocorrelation coefficient at lag k, $X_t$ is the activity level at time t, $Cov(X_t, X_{t-k})$ denotes the covariance between activity levels at time t and lagged time t-k, and $Var(X_t)$ represents the variance of the entire time series.

The ACF(k) value ranges between -1 and 1, where ACF(k) > 0 indicates a positive correlation in activity patterns at lag k, while ACF(k) < 0 suggests an inverse relationship. Higher ACF values indicate stronger temporal persistence in behavioral patterns, whereas a rapid decay in ACF suggests greater short-term fluctuations in activity.

## Support Vector Machine

XGBoost, an ensemble learning method based on gradient boosting, is well-suited for time-series classification tasks[25]. To train the model, we first organized the data into a feature matrix X and corresponding experimental condition labels y, where X represents the activity feature matrix of different individuals at various time points, and y denotes the experimental condition labels.

The classification model was trained using XGBoost, with Bayesian Optimization determining the max_depth parameter, while eval_metric='mlogloss' was specified to optimize for multiclass log loss, ensuring the model's adaptability to different experimental conditions. After training, we applied SHAP to analyze the feature contributions along the temporal dimension, identifying the most influential time points for classification[26–28].

SHAP values quantify the contribution of each time point to the classification outcome. To visualize the importance of features over time, we computed the mean absolute SHAP values across time points and examined their distribution. A SHAP threshold, defined as the mean plus 1.5 times the standard deviation, was used to identify the most significant time points influencing classification performance.

## Rhythmic Fluctuations Analysis

This study combines wavelet transform (WT) analysis and change point detection to explore the temporal dependence of Drosophila activity patterns and their rhythmic fluctuations[29]. Wavelet transform is a time-frequency analysis method that detects variations in signals across different time scales. Here, we used the Morlet wavelet as the mother wavelet function, which, as a Gaussian-modulated sinusoid, effectively balances time and frequency resolution, making it well-suited for biological rhythm analysis.

Wavelet transform provides a power spectrum across different time scales, enabling the assessment of rhythm intensity at various time intervals. To capture rhythmic patterns more precisely, particularly the 24-hour circadian cycle and longer periodic fluctuations, we employed a logarithmic scale distribution during wavelet transformation. We focused on temporal variations in wavelet power spectra and used the first-order derivative of power change to identify significant rhythm change points. Given P(t) as the average wavelet power at time t, rhythm change points were detected based on the time derivative:

$$\Delta P(t) = P(t+1) - P(t)$$

A time point was considered a potential change point if $|\Delta P(t)|$ exceeded a predefined threshold. To further enhance robustness, we integrated a peak detection method, identifying time points with the most pronounced power changes while setting a minimum interval (d) to prevent the detection of minor noise fluctuations.

Biologically, rhythm change points may correspond to adjustments in Drosophila activity patterns in response to environmental stimuli such as light, temperature fluctuations, or metabolic factors. For instance, if certain experimental conditions induce a significantly higher number of rhythm change points compared to controls, this may suggest an impact on the stability of the circadian clock. Additionally, this method allows for an in-depth investigation of the internal structure of activity patterns, revealing the presence of longer periodic rhythms beyond the circadian cycle, such as a potential 7-day rhythm or unstable rhythmic patterns.

## Principal Component Analysis

This study applied Principal Component Analysis (PCA) to reduce the dimensionality of Drosophila activity data, extracting major variance patterns while minimizing redundancy[30]. PCA identifies orthogonal directions that capture the greatest variance in the dataset by computing the covariance matrix and performing eigenvalue decomposition, projecting high-dimensional data into a lower-dimensional space.

We implemented PCA using the sklearn.decomposition.PCA module, selecting the first two principal components (n_components=2). All experimental group data were projected into a two-dimensional space, enabling the visualization of differences in activity patterns across experimental conditions.

## Similarity Score

To analyze Drosophila activity patterns under different experimental conditions, we trained a convolutional neural network (CNN) autoencoder to extract compact, high-dimensional representations of activity signals[31]. The autoencoder was designed to learn an efficient representation of continuous wavelet transform (CWT) images of activity signals while reconstructing the input as accurately as possible[32]. The encoder progressively reduces dimensionality, while the decoder reconstructs the original input.

During training, CWT-transformed activity signals were normalized between 0 and 1 to ensure model stability. The encoder consisted of two convolutional layers with max-pooling operations, enabling it to capture multi-scale temporal patterns, while the decoder employed upsampling and deconvolution layers to reconstruct the original signal. The model was optimized using the Adam optimizer with mean squared error (MSE) as the loss function to minimize reconstruction errors between input and output. After 50 training epochs, the model successfully learned key activity features, allowing for effective extraction of low-dimensional representations.

Following training, the encoder was used to extract latent features at each time point in the experimental data. Daily activity sequences were segmented into 24-hour windows, with each hour's activity signal transformed into a CWT image. These images were then passed through the trained CNN encoder, producing a low-dimensional feature vector for each hour:

To quantify the temporal similarity of activity patterns across experimental conditions, we constructed a reference dataset using elav-GS-GAL4 *Drosophila* recordings and computed the latent feature distributions for each hour. We used a weighted similarity metric based on Euclidean distance:

$$S(h) = \frac{\sum w_i e - dist(L_h, R_{h,i})}{\sum w_i}$$

where S(h) denotes the similarity score at hour h, $L_h$ is the latent feature for the experimental condition at that time, and $R_{h,i}$ are the reference latent features for the same hour.

## Plotting and t-testing

Plotting and mathematical calculations are performed by Python/R. Statistical significance testing was conducted using independent t-tests. Visualization was performed using ggplot(R), matplotlib and seaborn.

# Results

## No Causal Relationship between two periods of *Drosophila* Sleep

The fruit fly (*Drosophila melanogaster*), as a model organism, is widely used in circadian rhythm and behavioral studies. In this study, we analyzed fruit fly activity monitoring (DAM) data, standardizing hourly sleep in a numerical format and employing machine learning techniques to identify the influence of key temporal features. Specifically, we used an XGBoost classification model to distinguish activity patterns across different experimental groups and applied the SHAP method to interpret model decisions, thereby assessing the importance of temporal features. Furthermore, a time-series visualization based on SHAP importance scores revealed the contributions of different time intervals to classification decisions, enabling a quantitative evaluation of experimental variables on behavioral patterns.

This approach assumes that sleep patterns at any two hours are independent, meaning that sleep status at one time does not predict activity at another. Since daily rhythmicity operates autonomously, the duration of one sleep cycle does not significantly influence the next, at least in short-term monitoring. Moreover, the absence of hidden temporal dependencies ensures that sleep characteristics at each hour remain highly independent.

These assumptions ensure that any given sleep period constitutes an independent feature of fly sleep, similar to feature representations in classification models. This independence enables reliable scoring and comparison across different conditions.

To validate these assumptions, we utilized *elav-GS-GAL4*-driven fruit flies expressing RNA interference (RNAi) targeting *Drosophila* homolog of the monocarboxylate transporter 2 (*MCT2*), specifically the *UAS-sln RNAi* construct[33]. *GS-GAL4* is an inducible promoter activated by RU486, which, under the restriction of the neuron-specific *elav* promoter, drives RNAi expression in brain tissue[19]. By modulating the concentration of RU486, we were able to adjust *sln* expression levels, thereby progressively inhibiting ketone body uptake in neuronal bioactivity.

To examine correlations in sleep duration across a 24-hour period, we applied the SHAP method to analyze the Pearson correlation of features between different sleep intervals in the XGBoost classification model, generating a temporally structured heatmap. The results indicated no correlation between nighttime sleep episodes, whereas daytime sleep—particularly between 10 AM and 10 PM—exhibited noticeable associations (Figure 1A).

To further investigate the causal relationship between time-dependent features in *Drosophila* circadian behavior, we employed the DoWhy framework for causal inference, estimating the causal effects embedded within SHAP values. We implemented standardized preprocessing, backdoor adjustment regression, and statistical significance testing to construct and visualize the causal structure. This approach established a novel framework for time-dependent analysis, enabling the exploration of causal relationships within circadian behavioral dynamics. Our results were consistent with our expectations: while temporal correlations were present in the data, no causal relationships were detected among sleep features (Figure 1B). To evaluate the temporal rhythmicity of activity patterns, we then calculated the seven-day autocorrelation function (ACF). ACF quantifies the correlation between a time series and its lagged versions, providing insights into periodicity. The 24-hour sleep sequence in fruit flies exhibits short-term dependency, where recent hours have a stronger influence on the current sleep state, but this correlation gradually weakens over time. A periodic rhythm of approximately 12 hours emerges, and at longer time lags, a negative autocorrelation appears, potentially reflecting a compensatory fluctuation in sleep states (Figure 1C).

To further assess whether daily rhythmic patterns exhibit statistical independence across multiple days, we computed the Pearson correlation coefficients between diurnal cycle features over a seven-day period. The results indicated that rhythmic features between any two days within the seven-day sleep monitoring period showed no significant correlation, let alone a causal relationship (Figure 1D,E). A high autocorrelation at a seven-day lag suggests the presence of a weekly rhythmic pattern independent of circadian rhythms (Figure 1F).

## MCT2 in Nocturnal Sleep Stability and Circadian Transitions

To validate the value of our model and analytical pipeline in sleep research, we examined sleep homeostasis and circadian rhythms in Drosophila with progressively impaired MCT2 transporter function. Using our analytical framework, we investigated sleep-related changes to understand how the transport of lactate, pyruvate, and ketone bodies affects sleep timing. To assess the impact of MCT2 function on sleep architecture and circadian rhythms, we analyzed multiple sleep parameters in flies exposed to different concentrations of RU486 (0, 0.05, 0.1, and 0.5 mM). The sleep-wake cycle exhibited dose-dependent alterations. Control flies (0 mM) displayed a characteristic bimodal sleep pattern, with a primary sleep peak occurring early at night and a secondary peak in the late afternoon. However, as the concentration of RU486 increased, total sleep duration progressively decreased (Figure 2A,B), particularly during the nighttime. Statistical comparisons confirmed a significant reduction in nocturnal sleep duration at higher concentrations, indicating impaired sleep consolidation.

Sleep fragmentation was assessed by analyzing the distribution of sleep bout lengths and sleep episode counts (Figure 2C,D). Control flies exhibited prolonged and continuous sleep, whereas flies treated with RU486 showed a reduction in sleep bout length and a corresponding increase in sleep episode frequency, indicating increased sleep fragmentation. These effects were most pronounced at the highest concentration (0.5 mM), where sleep became highly fragmented, with almost no prolonged sleep episodes. Circadian rhythmicity was evaluated using locomotor activity records and wavelet analysis (Figure 2E,F). Control flies exhibited robust ~24-hour activity rhythms, whereas increasing concentrations of RU486 led to a progressive reduction in rhythm amplitude. At 0.5 mM, circadian rhythmicity was nearly abolished.

SHAP analysis of time-dependent features (Figure 2G) further identified critical time windows during which MCT function exerted the greatest influence on sleep patterns. The red dashed line denotes the threshold for peak definition, calculated as the mean + 1.5 × standard deviation. To account for potential confounding effects of total sleep duration, we normalized each fly's sleep duration using the mean of the activity time series. Previous studies have sporadically reported that ketone levels rise during sleep, suggesting a role in sleep maintenance. Our findings provide further insights into this process, demonstrating that ketone levels fluctuate and rise before and during the night, remaining elevated overall. Consistent with this, as we previously noted, ketones play a critical role at night, with our results revealing multiple peaks, suggesting a unique function in sleep maintenance and homeostasis. A novel observation from our study is that ketone levels also rise before awakening, a phenomenon not previously reported. We hypothesize that this pre-awakening increase, together with the pre-sleep peak, may regulate the sleep-wake cycle in Drosophila.

To further investigate the effects of MCT inhibition on circadian rhythms, we performed wavelet-based rhythm scoring in Drosophila (Figure 2H,I). Sleep-wake transitions were marked with dots, allowing us to track circadian rhythm changes over time. As concentration increased, these transition points became progressively fewer, indicating a weakening circadian rhythm. This reduction in phase transitions aligned with the previously observed dose-dependent disruption of sleep consolidation and increased sleep fragmentation. At the highest concentration (0.5 mM), transition points were nearly absent, reflecting severe disruption of the sleep-wake cycle. Additionally, with aging, the number of phase transitions continued to decline until they eventually disappeared. These findings suggest that the disruption of sleep homeostasis caused by impaired MCT2 function directly contributes to the progressive loss of circadian rhythmicity.

## Ketone Bodies and Sleep Disruptions in Alzheimer's Disease

We further investigated the effects and mechanisms of ketone body elevation on sleep in a Drosophila model of AD with metabolic dysfunction. To induce ketone body elevation, we employed two approaches. The first involved a soluble ketone salt formulation, allowing direct intestinal absorption of β-hydroxybutyrate (BHB), the primary ketone metabolite. The second approach relied on endogenous ketone induction through IF[34]. To control for potential confounding effects of increased sodium intake, we supplemented both the fasting medium and standard food of AD model flies, as well as the control diet, with equivalent concentrations of sodium chloride.

Control flies exhibited a severely disrupted sleep-wake cycle, characterized by fragmented and reduced sleep throughout the 24-hour period (Figure 3A,B). In contrast, both BHB and IF interventions significantly restored total sleep duration and continuity, particularly during the nighttime phase. Notably, statistical analysis revealed a significant increase in total sleep duration in the intervention groups compared to controls, suggesting a protective effect against APP+BACE1-induced sleep deficits.

Sleep fragmentation analysis further demonstrated that control flies experienced markedly reduced sleep duration and shorter sleep bouts (Figure 3C,D). Conversely, flies treated with BHB or IF exhibited increased sleep duration and fewer sleep episodes, indicating improved sleep consolidation. These effects were more pronounced in the BHB-treated group, implying a potential role of ketone metabolism in sleep homeostasis.

To assess the impact of BHB and IF on circadian rhythms, we analyzed multi-day locomotor activity patterns (Figure 3E,F). Control flies exhibited irregular and attenuated activity rhythms, with significantly reduced rhythmicity. In contrast, flies treated with BHB or IF displayed stronger and more stable circadian rhythms, as quantified by rhythm scores. Statistical comparisons confirmed a significant increase in rhythm scores in both intervention groups compared to controls, suggesting partial restoration of circadian function.

SHAP analysis (Figure 3G) identified the most pronounced intervention effects occurring in the two-hour period preceding night onset, further supporting the hypothesis that metabolic regulation can mitigate circadian and sleep disturbances in this AD model. However, during wakefulness (early light phase) and late-night periods, the ability of ketone bodies to sustain sleep homeostasis was limited.

Wavelet transformation analysis further evaluated whether ketone bodies facilitate wakefulness maintenance. Both BHB and IF significantly rescued sleep and wakefulness, with the observed effects progressively increasing over time (Figure 3H,I). This suggests that metabolic upregulation may have long-term therapeutic value in disease management.

## Ketone Dysregulation in Alzheimer's Disease Sleep Disruptions

Dysregulated lipid metabolism has been reported as a pathological feature of AD[35,36]. Previous studies have highlighted the critical role of ketone body deficiency in memory maintenance in energy-deficient animal models[37]. Based on this, we aimed to determine whether the sleep disturbances observed in AD could also be attributed to dysfunction in ketone-lactate exchange. To investigate this, we applied principal component analysis (PCA) to sleep pattern data under different conditions.

PCA of sleep patterns in sln RNAi knockdown flies revealed distinct clustering among treatment groups (Figure 4A). Control flies (0 mM, blue) exhibited a dispersed distribution, indicating high variability in sleep structure. In contrast, flies exposed to increasing concentrations of RU486 (0.05–0.5 mM) displayed a progressive shift in PCA space, with the highest dose (0.5 mM) forming a separate cluster, suggesting a strong regulatory effect on sleep architecture. A separate PCA analysis including APP+BACE1 expression and metabolic interventions (BHB and IF) revealed a similar trend (Figure 4B).

To isolate the influence of sleep duration on sleep pattern clustering, we normalized the activity time series mean. Under this condition, APP+BACE1-expressing control flies exhibited a sleep pattern model highly consistent with sln RNAi knockdown flies. However, flies treated with IF and BHB displayed a more discrete and differentiated sleep pattern, indicating that ketone body interventions altered sleep architecture

To systematically evaluate the effects of metabolic interventions on sleep structure and circadian rhythms, we constructed a convolutional autoencoder (CAE) to extract high-dimensional latent representations of sleep activity patterns. To assess the impact of metabolic interventions (BHB, IF) on circadian rhythms, we computed global similarity scores and hourly similarity curves, comparing experimental groups to a baseline reference model.

Based on RU486 dosage, fly phenotypes were categorized into a severity scale of 0–3, with 3 being the most severe. AD model flies were mapped onto this scale. Both BHB and IF treatments significantly reduced the severity score, indicating a functional rescue similar to sln RNAi reversal (Figure 4C). Hourly similarity analysis further reflected the ability of BHB and IF to modulate the same ketone transport pathway as sln RNAi. Notably, during light-to-dark transitions and pre-nighttime periods, BHB and IF treatment significantly reduced the severity score of AD flies, aligning with our previous SHAP analysis (Figure 4D).

This expansion of the SHAP findings suggests that while ketone bodies do not alter wakefulness patterns, they mitigate the inconsistency in wakefulness induced by ketone-lactate exchange dysfunction. More specifically, the inability of AD flies to initiate sleep is indeed ketone-related, but wakefulness itself is not entirely governed by ketone metabolism—at least, restoring ketone function alone does not alter it.

Furthermore, our analysis revealed that the nocturnal sleep homeostasis of AD flies appears largely independent of ketone metabolism and is not significantly influenced by ketone body elevation or depletion.

# Discussion

Sleep disorders such as insomnia, OSA, and narcolepsy often involve metabolic disruptions. Metabolomics research has identified biochemical signatures of these conditions, enabling personalized treatments[38–40]. Poor sleep is linked to metabolic syndromes like obesity, diabetes, and cardiovascular disease, with short sleep duration associated with insulin resistance, dyslipidemia, and inflammation[5,9,17,41–43]. These changes manifest in altered levels of IL-6, CRP, free fatty acids, and branched-chain amino acids, offering metabolic insights into sleep-related chronic diseases[44,45].

However, not all metabolites can be tracked non-invasively, and invasive sampling remains the standard. To address this limitation, this study introduces an approach that compares sleep data across metabolic states, revealing the temporal dynamics of metabolite function in sleep regulation.

## Temporal Dynamics of Sleep

Sleep is a complex physiological process involving dynamic changes across multiple levels, including brain activity, hormone secretion, metabolic regulation, and autonomic nervous system control[40]. While extensive research has identified patterns in these physiological parameters across sleep stages, their precise temporal relationships remain uncertain. Sleep is shaped by nonlinear and interactive processes; for example, melatonin secretion rises before sleep onset, while cortisol peaks in the morning, yet their exact interplay and impact on sleep quality are not fully understood[46,47]. The autonomic nervous system also exhibits stage-specific regulation—vagal activity dominates during non-rapid eye movement (NREM) sleep, whereas sympathetic activity is more pronounced during rapid eye movement (REM) sleep[48].

Sleep is not a single, continuous state but consists of multiple segments (e.g., nighttime awakenings, multi-phase naps). The relationship between these segments is complex. A notable example is the effect of naps on nighttime sleep: some individuals experience reduced sleep duration after napping, while others remain unaffected, possibly due to factors such as circadian rhythms and sleep pressure[49,50].

This study demonstrates that in a stable rhythmic framework, sleep episodes in flies are entirely independent of each other in terms of causality. However, this does not necessarily imply that all sleep episodes lack causal relationships. Our analysis focuses on the rhythmic fluctuations underlying sleep, using laboratory-bred flies with highly regular sleep patterns—conditions unlikely to occur in natural environments. Despite this limitation, our findings provide critical insights into the temporal organization of sleep. Under conditions free from environmental disturbances, sleep appears to be a rhythm-driven cyclic process rather than a highly autocorrelated event dependent on past states. The observed self-correlation between sleep episodes may, instead, reflect an adaptive response to environmental pressures. This suggests that sleep regulation may have evolved as a strategic mechanism to optimize energy allocation and enhance survival.

## MCT2 in Sleep Maintenance

This study used MCT2 to construct a model with impaired metabolite utilization. MCT2 regulates sleep by transporting ketone bodies and lactate, thereby influencing brain energy metabolism[33,51]. Ketone uptake promotes slow-wave sleep (SWS), while lactate metabolism affects GABA and adenosine levels, modulating sleep pressure and sleep architecture[51]. During sleep deprivation, MCT2 may enhance ketone and lactate utilization to facilitate restorative sleep. Our findings indicate that sleep in Drosophila is highly dependent on MCT2.

Further analysis reveals that MCT2 primarily affects sleep-wake transitions and nocturnal sleep homeostasis, likely mediated by the exchange and transport of lactate and ketone bodies. Metabolic studies suggest that lactate and ketone metabolism impact sleep structure and quality at different time points. Lactate levels typically rise during wakefulness and decline during sleep, reflecting shifts in brain metabolic activity[52]. Conversely, ketone bodies increase following sleep deprivation and enhance SWS under ketogenic conditions. In mice, six hours of sleep deprivation significantly elevates plasma ketone levels, accompanied by increased non-rapid eye movement (NREM) sleep, suggesting that ketone bodies contribute to sleep induction and maintenance[51,53]. When MCT2 function is inhibited, lactate and ketone metabolism become dysregulated, leading to the distinct SHAP curve variations observed at specific time points in our study.

## Ketone Bodies and Alzheimer's Disease

Research suggests that impaired glucose uptake is an early hallmark of AD. Monocarboxylate transporters (MCTs) facilitate the transport of ketone bodies and lactate in the brain and may play a crucial role in neuronal energy [54]. Although direct evidence of MCT2 dysfunction in AD is lacking, studies have reported impaired lactate metabolism and reduced expression of lactate transporters, including MCT2, in AD patients. This reduction may contribute to neuronal energy deficits, suggesting that MCT2 dysfunction could limit neuronal utilization of lactate and ketone bodies, exacerbating AD-related metabolic impairments[55].

Ketone bodies are considered an alternative energy source in AD. While glucose metabolism declines in early AD, the brain's ability to uptake and metabolize ketones appears preserved, making them a potential therapeutic option[56]. Clinical studies have shown that ketogenic diets or exogenous ketone supplements (e.g., medium-chain triglycerides, MCTs) enhance brain energy supply and improve cognitive function in AD patients[57,58]. In animal models, ketogenic diets have been shown to reduce Aβ and tau pathology while improving learning and memory in AD mice[59].

Our findings extend this understanding. The sleep patterns of AD flies closely resembled those with mild MCT2 suppression, shifting toward a more severe phenotype during circadian transitions. Ketone supplementation alleviated sleep fragmentation and circadian disruptions in AD flies. Moreover, ketone metabolism played a particularly prominent role during circadian transitions, highlighting its importance in maintaining rhythmic stability in AD.

Metabolic dysfunction accelerates AD progression through impaired glucose metabolism, mitochondrial dysfunction, and heightened inflammation, leading to Aβ accumulation and neuronal damage Improving sleep may help mitigate AD progression by stabilizing circadian rhythms and enhancing metabolic balance. Disruptions in glucose metabolism, mitochondrial function, and inflammation accelerate neurodegeneration, contributing to Aβ accumulation and neuronal damage[60–62]. Current therapies targeting circadian disturbances, such as light therapy, melatonin supplementation, and timed pharmacological interventions, have shown potential in improving both sleep quality and cognitive function[63,64]. Early intervention in circadian regulation may further slow disease progression, highlighting the importance of targeted strategies for neuroprotection and cognitive resilience[65]. Our approach provides additional evidence and new therapeutic possibilities for integrating sleep-based interventions into AD treatment.

# Data availability

All data and code supporting the findings of this study are available upon reasonable request from the corresponding author. Due to the nature of the study, data are not publicly accessible but can be provided upon request for academic and research purposes.

# Acknowledgements

This research was made possible through the invaluable support of the Adelaide Drosophila community, the Drosophila Facility at the University of Adelaide, and its outstanding researchers, Professor Robert Richards and Dr. Louise O'Keefe. We also extend our heartfelt condolences to memory of Dr. Louise O'Keefe.

This manuscript was edited using Overleaf, incorporating suggestions from language models, including Writefull's model and GPT-based models. AI-assisted tools were utilized to improve grammar, clarity, and academic style. All modifications were reviewed and accepted by the authors to ensure accuracy and adherence to the intended scientific meaning.

# Funding

No funding was received towards this work.

# Competing interests

The authors report no competing interests.

# Figure legends

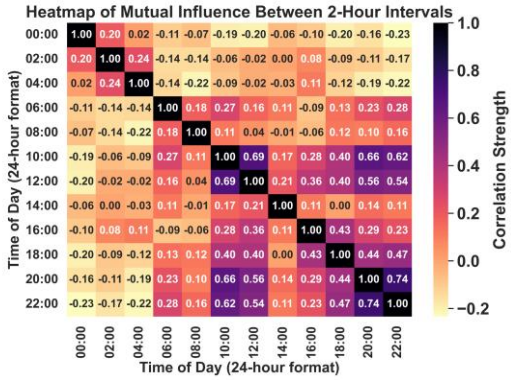
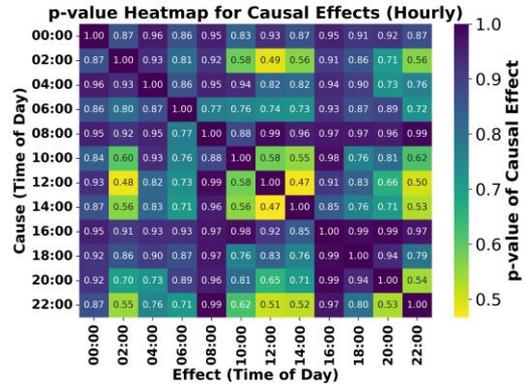
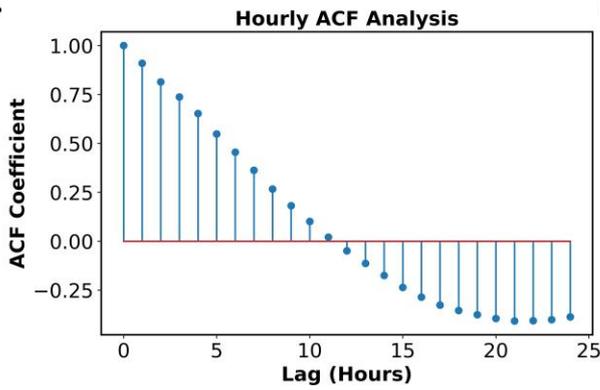
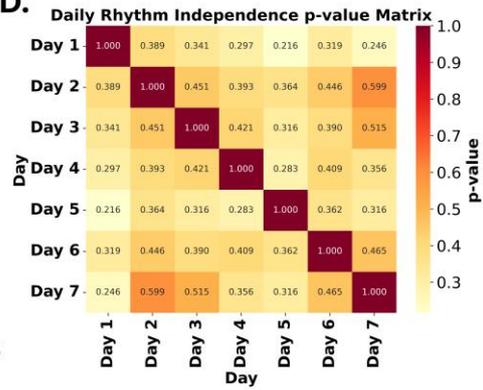
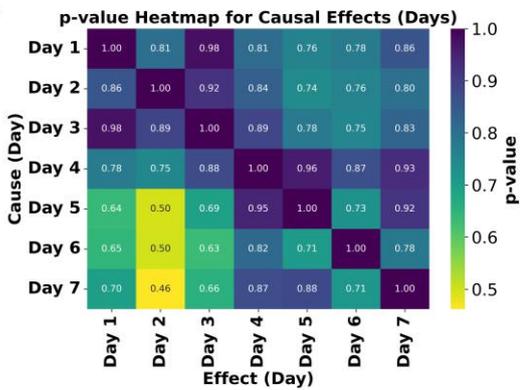
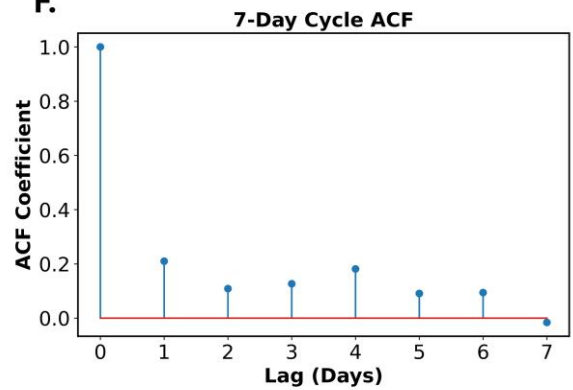

**Figure 1 Absence of Causal Link in Drosophila Sleep Timing.** (**A**) Heatmap of Pearson correlation illustrating mutual influences between 2-hour intervals over a 24-hour cycle. Data were processed by resampling original feature dimensions into 30mins intervals, then aggregated into 2-hour bins to compute Pearson correlation coefficients. (**B**) Hourly heatmap of p-values for inferred causal effects, derived from DoWhy causality tests with a maximum lag of 3 hours, highlighting statistically significant temporal causation. (**C**) ACF analysis at hourly resolution, demonstrating cyclical decay in correlation over a 24-hour lag. (**D**) p-value matrix assessing the independence of daily rhythms across seven consecutive days, where lower values indicate stronger evidence against independence. (**E**) Daily causal effect heatmap displaying the statistical significance of day-to-day causal relationships over seven days. (**F**) Seven-day cycle ACF analysis, showing a strong correlation at a one-day lag, followed by a rapid decline, reinforcing the presence of a daily rhythmic pattern.

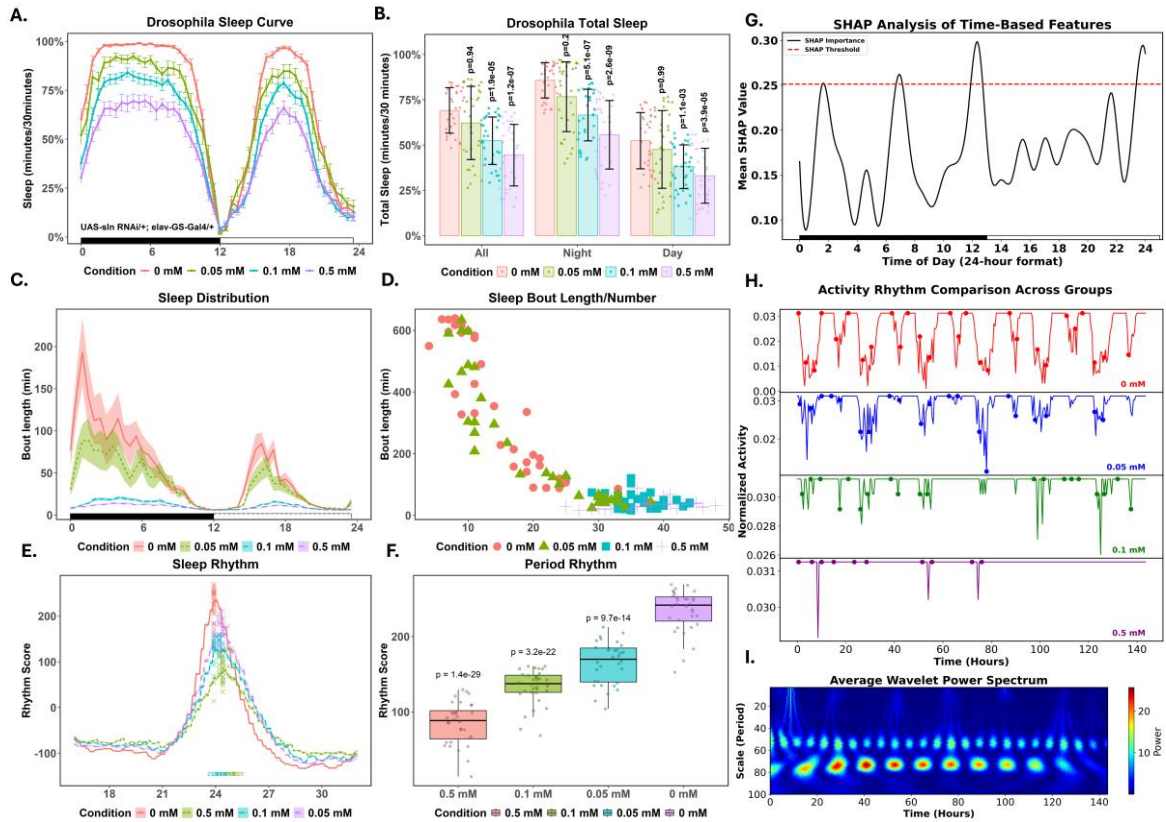

**Figure 2 Sleep and Activity Patterns Under Varying RU486 Treatment Conditions. (A)** Average sleep (minutes per 30-minute intervals) over a 24-hour cycle in **Drosophila** at RU486 concentrations of 0 mM, 0.05 mM, 0.1 mM, and 0.5 mM, revealing distinct diurnal sleep patterns. **(B)** Total sleep duration (minutes) across all time points, nighttime, and daytime intervals for each RU486 concentration, with statistical significance indicated. **(C)** Sleep bout length distribution over 24 hours for each RU486 concentration, highlighting differences in bout duration patterns. **(D)** Sleep bout length distribution over 24 hours for each RU486 concentration, highlighting differences in bout duration patterns. **(E)** Averaged sleep rhythm across RU486 concentrations, showing peak rhythmicity around midday. **(F)** Box plot of period rhythm scores across RU486 concentrations, indicating significant variability in rhythmicity. **(G)** SHAP value analysis identifying influential time-based features in model predictions, with the significance threshold set at mean + 1.5 × standard deviation. **(H)** Normalized activity rhythms across different RU486 treatment groups, highlighting treatment-specific temporal activity patterns. **(I)** Continuous wavelet transform-based average power spectrum visualizing rhythmic activity patterns, demonstrating distinct rhythmic power densities over time. n=32 per condition. Statistical significance was assessed using Welch's t-test.

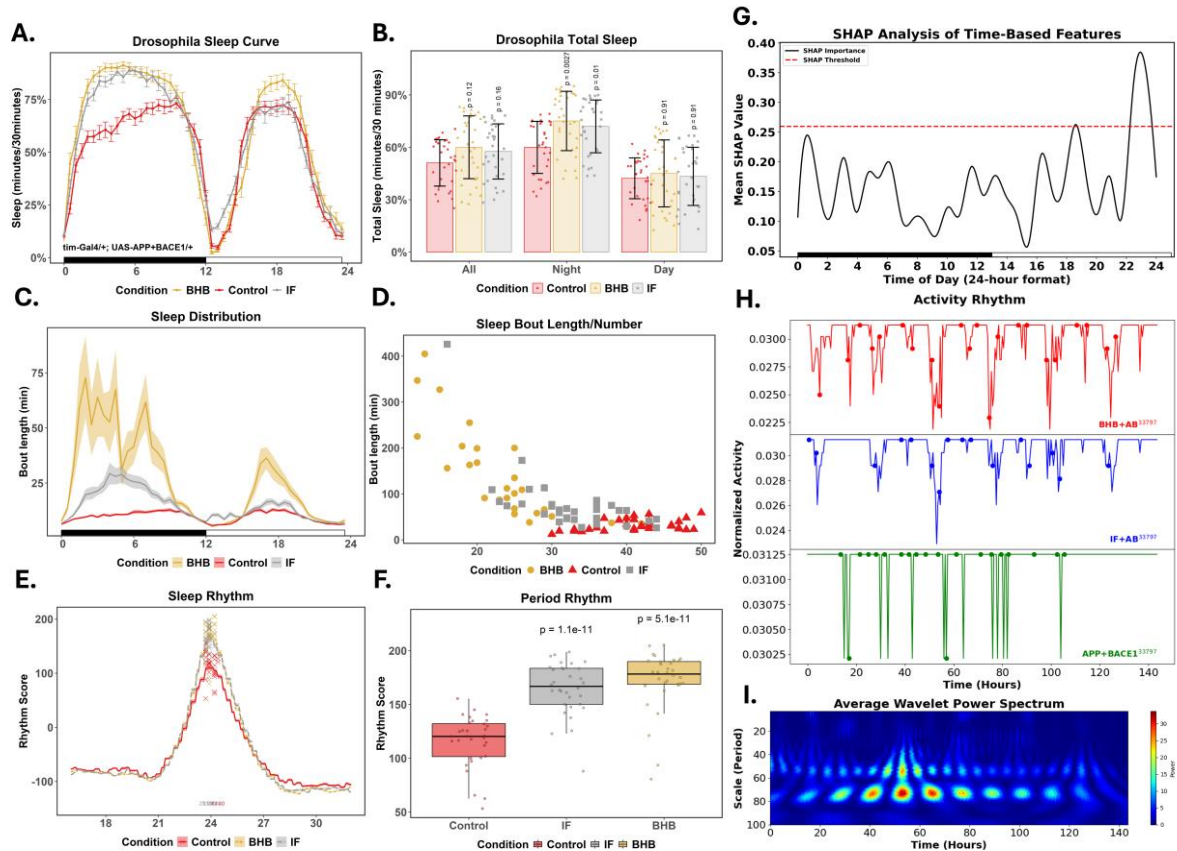

**Figure 3 Sleep and Activity Analyses Under BHB Treatment and Intermittent Fasting in Drosophila Expressing APP+BACE.** (**A**) Average sleep (minutes per 30-minute intervals) over a 24-hour cycle comparing BHB treatment, IF, and control groups, highlighting altered diurnal sleep patterns in AD flies. (**B**) Total sleep duration across all, nighttime, and daytime periods, with statistical differences indicated. (**C**) Sleep bout length distribution over 24 hours for each condition, emphasizing variations in bout duration patterns. (**D**) Scatter plot of sleep bout length versus the number of bouts, illustrating distinct sleep architectures across conditions. (**E**) Averaged sleep rhythm score for each condition, revealing differences in rhythmicity, particularly around the daily peak. (**F**) Box plot of rhythm scores across control, IF, and BHB groups, indicating significant variability. (**G**) SHAP analysis of time-based features influencing model predictions, with the significance threshold set at mean + 1.5 × standard deviation. (**H**) Normalized activity rhythm comparisons across groups, demonstrating treatment-specific temporal activity patterns. (**I**) Continuous wavelet transform-based average power spectrum visualizing rhythmic activity patterns, highlighting distinct rhythmic power densities across conditions. n=32 per condition. Statistical significance was assessed using Welch's t-test.

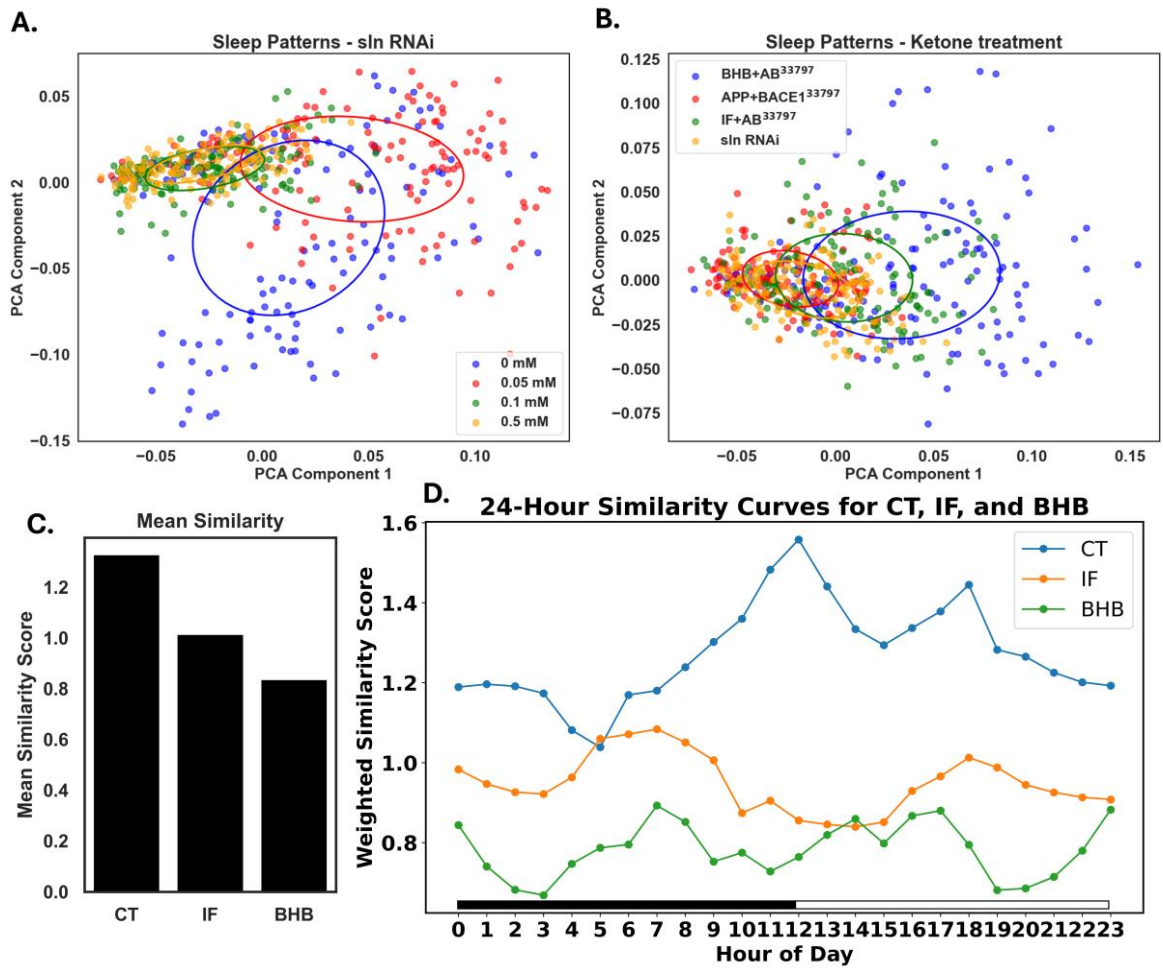

**Figure 4 PCA-Based Clustering and Similarity Analyses of Sleep Patterns Under RNAi and Treatment Conditions.** (**A**) PCA scatter plot illustrating sleep patterns in **Drosophila** expressing *sln* RNAi under varying RU486 concentrations (0 mM, 0.05 mM, 0.1 mM, and 0.5 mM), driven by *elav-GS-GAL4*, demonstrating condition-specific clustering. (**B**) PCA scatter plot comparing sleep patterns across groups: BHB + AB, IF + AB, *APP+BACE1* driven by *tim-GAL4*, and *elav-GS-GAL4 > sln* RNAi under 0.1 mM induction, with 95% confidence ellipses. (**C**) Mean similarity scores computed via autoencoder-based latent feature extraction, comparing control, IF, and BHB conditions. Lower scores indicate greater divergence from control rhythms. (**D**) Smoothed 24-hour weighted similarity curves derived from latent representations generated by convolutional autoencoder modeling, displaying hourly similarity dynamics for control, IF, and BHB treatment groups.